\documentclass{amsart}
\usepackage[T1]{fontenc}
\usepackage[latin1]{inputenc}
\usepackage{graphics}
\usepackage{setspace}
\makeatletter

\providecommand{\LyX}{L\kern-.1667em\lower.25em\hbox{Y}\kern-.125emX\@}

 \theoremstyle{plain}    
 \newtheorem{thm}{Theorem} 
 \theoremstyle{plain}    
 \newtheorem*{thm*}{Theorem} 
 \theoremstyle{plain}    
 \newtheorem{cor}{Corollary} 
 \theoremstyle{plain}    
 \newtheorem*{cor*}{Corollary}
 \theoremstyle{plain}    
 \newtheorem{lem}{Lemma} 
 \theoremstyle{definition}
 \newtheorem{defn}{Definition}
 \theoremstyle{definition}
 \newtheorem*{defn*}{Definition}
 \theoremstyle{remark}    
 \newtheorem*{acknowledgement*}{Acknowledgement} 

\makeatother

\begin{document}

\title[Pricing Virtual Paths with QoS Guarantees as Bundle Derivatives]{Pricing Virtual Paths with Quality-of-Service Guarantees as Bundle Derivatives}

\author{Lars Rasmusson }

\date{2001-06-12}

\address{Swedish Inst. of Computer Science \\
Box 1263, S-16429 Kista, Sweden}

\email{\texttt{Lars.Rasmusson@SICS.se} }

\begin{abstract}
We describe a model of a communication network that allows us to price complex
network services as financial derivative contracts based on the spot price of
the capacity in individual routers. We prove a theorem of a Girsanov transform
that is useful for pricing linear derivatives on underlying assets, which can
be used to price many complex network services, and it is used to price an option
that gives access to one of several virtual channels between two network nodes,
during a specified future time interval. We give the continuous time hedging
strategy, for which the option price is independent of the service providers
attitude towards risk. The option price contains the density function of a sum
of lognormal variables, which has to be evaluated numerically. 
\end{abstract}
\maketitle

\section{Introduction}

\subsection{End-to-end quality of service }

Today, most traffic in computer networks is handled by best effort routing;
each network router passes on packets as long as it can, and when the buffers
are full, it drops incoming packets. When the network load is low, all data
streams get a high throughput, and when the load is high, all streams experience
equal loss.

This works well for some data streams such as file transfer, but less so for
real-time data streams, i.e. when data packets have hard deadlines. Examples
are audio/video streams \cite{faratin00automated}, grid computing\cite{krauter00architecture},
and interactive data streams. Congestion causes packet losses and retransmissions
that result in jitter, suspended computation, and high latency, respectively.
These problems arise because the network cannot provide service quality guarantees
and different service levels for different kinds of traffic. Ability to provide
service guarantees requires that an individual user can reserve some of the
capacity in the congestion prone parts of the network, be that routers, network
links, or whatever.

The flexibility of today's computer network is due to the design choice to keep
the logic inside the network very simple, and to let all application specific
knowledge be handled {}``at the ends{}'', by applications on top of the network
layer. In this spirit, we advocate that a reservation policy should be implemented
outside of the network, and that the network should only be a delivery vehicle
for packets. Current attempts to improve throughput rely on more complex internal
statistical routing and network maintenance. This {}``intelligent network{}''
principle is different to the {}``end-to-end principle{}'', and intelligent
networks have not been as good as end-to-end networks at supporting new applications
and uses of the network.

\subsection{End-user bandwidth markets}

In today's parlance, discussions of of bandwidth markets often refer to the
trading of spare trunk capacity among large telecom companies, Internet service
providers, etc. See for instance the bandwidth markets at Band-X, RateXchange,
Min-X, etc. In these markets, the purpose of trading is to maximize the profit
of the service providers, i.e. to let them fulfill their prior client obligations
at a low cost. Since end-users are not affected by the cost, these prices only
affect the traffic load on a coarse scale. Service providers can only guess
what the best buy would be, since they do not know the network requirements
of the applications running on the end-users' computers.

We propose a somewhat different approach. To make an efficient market, the bandwidth
allocation decisions should be a fine grained negotiation about access to the
scarce resources, that takes place between the end-users, the actual consumers.
This way, someone that really needs a particular resource can bid for it high
enough that someone else releases (sells) the resource, and buys another resource
instead. In most cases, end-users require complex services, services that require
capacity in more than one router, and where these prices interact in a complex
way to form the total price of the service.

\begin{figure}
{\par\centering \resizebox*{!}{3cm}{\includegraphics{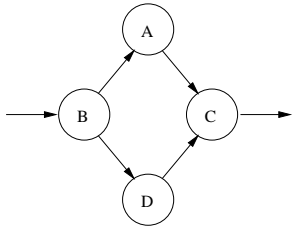}} \par}

\caption{\label{fig:net1}An end-user wishes to reserve capacity in a path from B to
C, i.e. buy \{B,A,C\} or \{B,D,C\} or neither one, if the price is too high.
The total price depends on the prices of the individual resources in a complex
way.}
\end{figure}
 
\begin{figure}
{\par\centering \resizebox*{!}{4.2cm}{\includegraphics{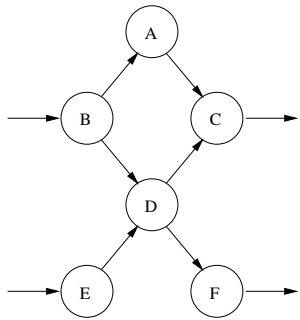}} \par}

\caption{\label{fig:net2}User 1 has reserved a path along \{B,D,C\}, and user 2 wants
to reserve \{E,D,F\} but the price of D makes the path too expensive. Then,
if the prices are appropriate, user 1 should sell D and buy A, and user 2 buys
D.}
\end{figure}

The resource prices are correlated in many ways. In fig. \ref{fig:net1}, the
prices of all resources that are on a potential path affect the decision on
whether do buy any of the other resources. In fig. \ref{fig:net2}, the demand
of user 2 affects the price of resource A, \emph{not} on the user 2's potential
paths. 

There are some hurdles to overcome to make the resource negotiation fast and
efficient:

\begin{enumerate}
\item The negotiation between the end-users must be kept very simple. Bilateral negotiations
\cite{faratin00automated} is infeasible in real applications.
\item An end-user does not get a definite price-quote for a complex service such as
a path that involves capacity in several routers.
\item The end-users will generate an extreme amount of network traffic when they buy
and sell resources, get price quotes, etc.
\end{enumerate}
These problems are addressed by the method presented here.

\subsection{Related work}

Related work on bandwidth markets to improve performance in networks are generally
based on admission control at the edges, as done by Gibbens \emph{et al.} \cite{gibbens99resource}.
A user is not admitted to the network if the network does not provide sufficient
service quality. For instance, Kelly \cite{kelly01mathematical} models interconnections
as reservations along a specified path, and gives Lyapunov functions that show
that the system state stabilizes asymptotically, as end-users change their demand
according to network load. Courcoubetis \cite{courcoubetis01providing} models
a router as consisting of \( C \) channels, derives the probability that a
certain fraction of the traffic is lost, and prices a call option that bounds
the price a user has to pay for capacity in one router. Semret \emph{et al}.
\cite{semret99spot} model admission control to a network over an exponential
distributed number of minutes as a number of n:th-price auctions on 1 minute
time-slot access, and price an access option as the sum of call options on each
of the time slots. Lukose, \emph{et al}. \cite{lukose97methodology} use a CAPM-like
model to construct a mixed portfolio of network access with different service
classes, in order to reduce the latency variance and mean.

The above models either investigate the asymptotic network behavior, or only
model the price for one network resource. We are interested in handling the
transient behavior of a network with several interacting nodes. 

In neither of the models above do the underlying resources constitute a complete
market, i.e. it is not possible to create a momentarily perfectly balanced (risk-less)
portfolio of options and underlying resources. The price of the option is therefore
dependent on the risk-aversiveness of the network provider. In a complete market,
the price is independent of the risk-aversiveness since perfect hedges can be
created, and the price can be set more tightly, since anyone can trade and compete
for the bids. 

We will present a model of a complete continuous time market that allows us
to derive the price of an option that extends the capabilities of the above
options in several ways 

\begin{itemize}
\item the price depends on more than one underlying resources
\item the actual path, or set of resources assigned, does not have to be specified
in advance
\item the price is risk-neutral, using the Black-Scholes' assumptions
\end{itemize}
Furthermore, we choose a market type in which the resources are traded continuously,
rather than in auctions with discrete clearing times since that causes latency.

\subsection{Structure of the paper}

In section \ref{sec:finance} we recapture relevant formulas from the theory
of derivative pricing, in sec. \ref{sec:model} we describe the model price
process for which we price the derivatives, and models of the market and network
resources. In sec. \ref{sec:pricing} we state the main results, which are the
definitions and price formulas for the network option (proofs are deferred to
the appendix), and we conclude with a discussion in sec. \ref{sec:discussion}.

\section{Preliminaries\label{sec:finance}}

\subsection{Derivative pricing}

A standard way to price derivative contracts, a.k.a {}``derivatives{}'', such
as options, futures, etc. is to use arbitrage-free portfolio theory, which says
that an asset, known to be worth \( S_{T} \) at a future time \( T \), is
worth \( S_{0}=e^{-rT}S_{T} \) today. Here \( r \) is the continuously compounded
interest {}``risk-free rate{}'', the loan/interest rate that you get from
a bank or a government bond. The reasoning is that if the asset was worth \( X\neq S_{0} \),
which is less (more) than \( S_{0} \), then anyone could make money by borrowing
(lending) \( X \) to the rate \( r \), buy (short sell) the resource, wait
to \( T, \) sell (buy back) the resource for \( S_{T} \), pay back \( Xe^{rT} \)
(withdraw \( X^{rT} \)) and keep the arbitrage profit \( S_{T}-Xe^{rT}>0 \)
(\( Xe^{rT}-S_{T}>0 \)). This cannot be possible in an arbitrage-free market.
The argument requires that short selling is allowed, and that transaction fees
are negligible. 

Derivative pricing often models the asset prices as It\^o processes \( S(t) \),
\[
dS(t)=a(t,S(t))dt+b(t,S(t))dW(t)\]
 where \( W(t) \) is a Wiener process, and \( a \) and \( b \) are sufficiently
bounded functions \cite{kloeden99numerical}. The Black-Scholes method \cite{black73pricing}
prices a derivative of an asset which price follows a particular kind of It\^o-process.
It is based on constructing a portfolio that invests some part of its money
in the option, and some part in the asset. At each instant, the portfolio is
balanced in such a way that its value after \( dt \) is known exactly. As time
goes and prices change, the portfolio is rebalanced. In short, it is shown that
a derivative \( f(t,s) \) that is a function of a stochastic process\[
dS(t)=a(t,S(t))dt+\sigma S(t)dW(t)\]
 follows the Black-Scholes equation\[
\frac{\partial f}{\partial t}(t,s)+rs\frac{\partial f}{\partial s}(t,s)+\frac{\sigma ^{2}s^{2}}{2}\frac{\partial ^{2}f}{\partial t^{2}}(t,s)-rf(t,s)=0\]
\[
f(T,s)=g(s)\]
which from the Feynman-Kac formula can be seen to have the solution \[
f(t,s)=e^{-r(T-t)}E^{Q}[g(S(T))|S(t)=s]\]
 where \( Q \) is the so called equivalent, or risk-free, martingale measure.
The boundary condition, given by the function \( g(s) \), specifies the value
of the option at the time the derivative expires. For a traditional call option,
\( g(s)=max(s-K,0) \), where \( K \) is the strike price. 

Under the \( Q \) measure, \( S(T) \) has the drift term \( a(t,s)=rs \),
and hence, under \( Q \), \[
S(T)=S(t)e^{(r-\frac{\sigma ^{2}}{2})(T-t)+\sigma (W(T)-W(t))}\]
Recall from the definition of It\^o processes that \( W(t) \) is a Wiener process,
in other words, \( W(T) \) is normal distributed with mean 0 and variance \( T \)
given \( F_{0} \), the knowledge up until \( t=0 \). This means that we can
price derivatives using Monte-Carlo simulation to solve the differential equations
above. Another advantage with Monte-Carlo is that it converges quite fast also
for multidimensional problems, something that is not the case for binomial-tree
pricing methods.

Rebalancing a portfolio, or obtaining/selling assets to decrease its variance
is called 'hedging' the portfolio. The Black-Scholes hedging method produces
a self-financing hedge, i.e. no additional capital is needed to balance the
risk of the portfolio. Another interesting effect of Black-Scholes hedging is
that the formula for the optimal continuous-time hedge at \( t \) given \( S(t) \)
does not involve the drift function \( a(t,s) \). Hence, the continuous-time
hedging is the same for the mean-reverting and the exponential drift processes.

\subsection{Applied pricing}

The assumption that a portfolio can be continuously rebalanced is violated in
real markets. Market frictions, such as transaction fees, often make it too
costly to rebalance a portfolio very often. For bandwidth markets, we can eliminate
market friction all together, since we are free to design the market to our
own liking. We cannot however guarantee that we can rebalance the portfolio
at every instant, since there are others that want to trade concurrently with
ourselves, and multiple other trade events may occur between the rehedging events.

To understand the effect of interval hedging compared to continuous time hedging,
we show the effect on the portfolio value of hedging and rehedging a call option
on a single asset for three different price processes, hedged continuously and
at intervals.

\subsubsection{Continuous time hedging}

The lognormal Brownian motion process is often used to model stock stock price
\( S(t) \). Its dynamics is \begin{equation}
\label{eq:lognorm}
dS(t)=\mu S(t)dt+\sigma S(t)dW
\end{equation}
 A derivative contract, on an underlying asset obeying (\ref{eq:lognorm}) and
\( \mu =r \) under \( Q \), can be hedged in continuous time by creating a
portfolio of \( \gamma (t) \) derivatives and \( \beta _{i}(t) \) assets.
A perfectly hedged, self financing, portfolio with a derivative contract depending
on \( N \) assets \( \bar{S}(t)=\{S_{1}(t),...,S_{n}(t)\} \) has the value
\[
\Pi (t)=\gamma (t)\, f(t,S(t))+\sum _{i=1}^{N}\beta _{i}(t)\, S_{i}(t)\]
 follows (for lack-of-arbitrage reasons) the value of a safe investment, \( \Pi (t)=\Pi (0)e^{rt} \),
where \( r \) is the risk-free continuously compounded interest rate. The hedging
strategy is \begin{eqnarray}
\gamma (t+dt) & = & \frac{\Pi (t)}{f(t)-\sum _{i=1}^{N}\frac{\partial f}{\partial S}(t)\, S(t)}\label{gamma} \\
\beta _{i}(t+dt) & = & -\gamma (t)\frac{\partial f}{\partial S_{i}}(t)\label{beta} 
\end{eqnarray}

\vspace{0.3cm}
{
\begin{figure}
{\par\centering \resizebox*{0.7\textwidth}{!}{\rotatebox{90}{\includegraphics{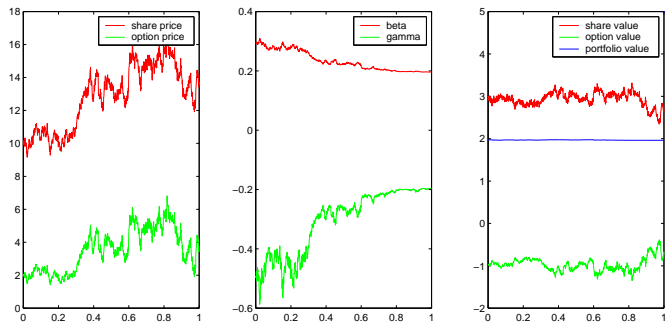}}} \par}

\caption{To the left, the value of an asset and a call option with strike price \protect\( K=10\protect \),
in the middle, the fraction invested in each resource to get a balanced portfolio,
and to the right, the total value of the portfolio (constant), and its parts.\label{fig:hedge1}}
\end{figure}
\par}
\vspace{0.3cm}

Consider a share whose price follows a lognormal Brownian motion, see fig. \ref{fig:hedge1}.
The price is plotted in the left graph, together with the price of a call option
with strike price 10 that expires at \( t=1. \) The middle graph shows the
composition of a perfectly hedged portfolio. The topmost curve is \( \beta (t) \),
the parts of the portfolio part invested in shares, and under it is \( \gamma (t) \),
the part invested in options. A negative number means that the option should
be sold short. To the right is a plot of the total value of the portfolio that
initially consisted of one option. Above and below are plots of the total value
of the portfolio holdings, in shares and options, respectively. Here \( r=0 \),
so the value of the portfolio should be constant even though the share and option
prices fluctuate. Since the portfolio value is constant in the rightmost plot,
the hedging works, and the portfolio yields the risk-free rate.

A well known result of the Black-Scholes pricing is the somewhat surprising
result that the derivative prices is not dependent of the drift term. This is
because it is based on a first order approximation, and the drift term is \( O(dt) \)
while the diffusion term is \( O(\sqrt{dt}) \). Since the drift term is the
only difference between the lognormal process and the mean-reverting process
with multiplicative noise, the hedging scheme works as equally well for both
processes.

\subsubsection{Interval time hedging}

\begin{figure}
{\par\centering \resizebox*{0.7\textwidth}{4cm}{\rotatebox{90}{\includegraphics{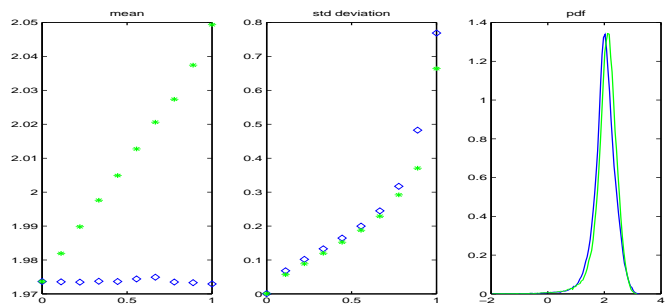}}} \par}

\caption{To the left and in the middle are plots of average value and standard deviation
for a portfolio at the rehedging times. To the right is the portfolio value
distribution when the option expires, \protect\( (t=1)\protect \). Derived
numerically (by Monte-Carlo simulation) for a lognormal price process (diamonds),
and for a mean-reverting process (stars).\label{fig:fig4}}
\end{figure}
For hedging at discrete events with an interval \( \Delta t>0 \) rather than
in continuous time, the above formula does not give a complete hedge. 

The effect of hedging a call option with an unmodified strategy is shown, for
two different underlying processes, in fig. \ref{fig:fig4}. Above to the left,
is a plot of the average portfolio value from a Monte-Carlo simulation, of a
lognormal process, and a mean-reverting process with multiplicative noise. The
middle plot shows the variance of the portfolio value at the 10 rebalancing
times. The variance increases with time, and the variance is higher for the
mean-reverting process than for the lognormal process. To the right is the density
function for the portfolio value at \( t=1 \). The rebalancing of the portfolio
is only done every 100 steps, i.e. \( \Delta t=0.1 \). It is apparent that
the portfolio value for the mean-reverting process is not constant. It is hence
possible to make a statistical arbitrage profit on the derivatives. However,
the deviation in expected value is only a few percent, so a small increase in
the derivative price can protect the issuer from the arbitrage risks.

For some processes a modified hedging strategy can be derived. Cornalba, Bouchaud
\textit{et al}. \cite{bouchaud00option} have considered time correlated stochastic
processes and shown that for the Ornstein-Uhlenbeck process \[
dS(t)=\alpha (\mu -S(t))dt+\sigma dW(t)\]
 the same hedging strategy can be used, but with a modified variance. A similar
derivation, inspired by Bouchaud, gives that \( \hat{\sigma }^{2}=\sigma ^{2}\frac{(1-e^{-2\alpha \Delta t})}{2\alpha \Delta t} \).
This is based on a first order approximation, hence the modified volatility
\( \hat{\sigma } \) is appropriate only for small \( \Delta t \). For many
processes, such as processes with multiplicative noise, we do not have modified
strategies.

\begin{figure}
{\par\centering \resizebox*{0.7\textwidth}{4cm}{\rotatebox{90}{\includegraphics{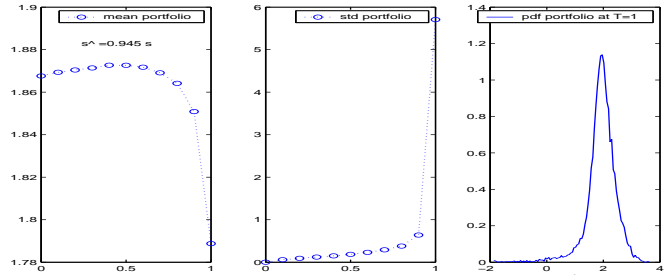}}} \par}

\caption{Plots of mean, standard deviation, and final distribution of portfolio value
(see fig. \ref{fig:fig4}) for a mean-reverting asset price process, rehedged
10 times with the adjusted variance \protect\( \hat{\sigma }^{2}\protect \).
\label{fig:fig5}}
\end{figure}
Fig. \ref{fig:fig5} shows plots, similar to fig. \ref{fig:fig4}, of the values
for a mean-reverting process that is hedged with the modified measure of variance
\( \hat{\sigma }^{2} \), and it can be seen that the adjustment is not perfect,
but still only a few percent off. Its dynamics are more complex, and we do not
know of a strategy with modified variance that makes the portfolio value process
replicate that of a risk-free portfolio. 

Since the portfolio cannot be made risk-free over all of \( \Delta t \), the
future portfolio value is uncertain. In terms of mathematical finance, the market
is incomplete. In complete markets, options can be priced in a way that does
not depend on the risk-aversiveness of individual participants, something that
is not the case in an incomplete market. However in a technical system, designed
to be controlled by a market, we can require there to always be one or more
trading programs that behave risk-neutrally, or that charge a specified risk
premium. That guarantees that derivatives are traded at prices that make the
market efficient, in the sense that they maximize the expected utility of the
end-users by providing low option prices.

\section{Model\label{sec:model}}

\subsection{Price processes}

In a computer network, end-users share the limited capacity of the routers and
links. To be able to provide QoS-dependent services, these resources must be
managed by the end-users. To each end-user, the system load appears to fluctuate
stochastically, as the end-user does not have access to the complete system
state. 

Our approach is to view the state of the system as stochastic processes, and
to control the use of scarce resources by trading the usage right at spot markets.
On these markets, prices will appear to be stochastic processes. The prices
present an an aggregated view of the system. With this view, controlling a technical
system with interacting subunits (not necessarily a network), boils down to
developing suitable derivatives and hedging schemes for the different services
that the system should provide. The implementation requires market-places for
the individual resources, and third party middle-men that sell derivatives to
end-users and do the actual trading on the resource level. 

An average, sporadic end-user is not willing to take the risk of paying an excessive
amount for a network service, but rather get a definite price quote. Trading
derivative contracts is a trade of \emph{risk,} where one part gets the risk
and a premium, and the other part gets a fixed cash flow. With suitable market
models, derivative contracts can be priced objectively, at least when the price
processes can be described sufficiently well. So, instead of trading the actual
resources, the end-users buy derivative contracts of a third party. The contract
guarantees the delivery of the required set of resources. The contract may specify
both future delivery of some resources, and deliveries that are contingent on
future prices, or functions thereof. 

In a previous paper \cite{rasmusson01price}, we have simulated a simple bandwidth
market without derivatives, to determine the properties of the resulting stochastic
price process. It was found to be very well described by a correlated mean-reverting
process with multiplicative noise, \begin{equation}
\label{eq:meanrev}
dS_{i}(t)=\alpha _{i}(\mu _{i}-S_{i}(t))dt+\sigma _{i}S_{i}(t)dW_{i}(t)
\end{equation}
where the price correlation \( Corr[dW_{i}(t),dW_{j}(t)]=\rho _{ij} \). We
gave estimations of the parameters \( \alpha ,\mu ,\sigma ,\rho , \) based
on price history data from the simulation. Since this modeling was possible,
it appears feasible to represent complex network services in terms of derivative
contracts on certain complex combinations of resources. Since in general, introducing
new assets in a market modifies the price dynamics, hence the parameters must
be re-estimated when new derivatives are added.

The mean-reverting process drifts back towards \( \mu  \) with a rate determined
by \( \alpha  \). As opposed to lognormal processes, mean-reverting processes
are auto-correlated processes, and their the variance per time unit, \( \frac{1}{\tau }Var[S(t+\tau )-S(t)] \)
decreases with increasing \( \tau  \) for the mean-reverting process, while
it is constant for the lognormal process. It is the fact that the price process
has {}``memory{}'' that causes the deviation in the expected value of the
portfolio for interval hedging.

\subsection{Farmer markets}

The market places where resource trading are of a special kind designed for
very rapid markets that we call Farmer markets, as the original inspiration
was found in \cite{farmer00market}. Each market handles one resource, and is
run by a market maker that at each instance guarantees to accept bids both to
buy and to sell. 

There is no back-log of pending {}``limit orders{}'', only bids {}``at market{}''
are accepted. This guarantees that the trading can take place with very little
overhead for the market maker. Since only bids at-market are accepted, the bidder
does not know at what price resources will be traded, but prices can be estimated
from the price-quote history. 

The central idea of this market design is that the resources are exchanged on
these markets, and that all more complex contracts are expressed as derivative
contract on these resources. For instance, a limit order, i.e. an order to buy
if the price is less than a specified amount, is a risky contract since the
bidder does not know if the deal will go through or not.

\subsection{Resource shares}

The capacity of each resource is divided into equal well defined shares, that
gives the owner the right to send a certain amount of traffic on a short time
\( \Delta t \) if he pays \( \epsilon \, S(t)\Delta t \). From here on, we
assume \( \epsilon =1 \). The total capacity of the shares must not surpass
the total capacity of the resource. Without the payment from the resource holder
to the market maker, a holder of the resource would have no incentive to avoid
congested resources, which is shown in the pricing of the bundle options below.

For routers, statistical multiplexing has shown to give a large throughput increase,
hence it seems reasonable to mix two traffic classes. A router has two traffic
classes, 1) the guaranteed class, and 2) the best-effort class. Packets in the
first class are guaranteed not to be dropped in case of congestion, while packets
without valid credentials are handled in traditional best-effort manner. As
with the Metro Pricing Scheme by Odlyzko \cite{odlyzko99paris}, we only requires
two traffic classes, but in the Metro Pricing scheme, prices are determined
outside of the system in such a way that there is no congestion in the first
class, while in our model, prices are determined by demand, and first class
packets get to go first if there is congestion.

\section{Results\label{sec:pricing}}

In a computer network, end-users want to establish virtual paths with certain
guaranteed properties, such as loss, latency, etc. The user wants to have the
resources at \( T_{1} \) and to sell them again at \( T_{2} \). This can be
implemented as an option that delivers the resources at \( T_{1} \) together
with a bundle of options that lets the user sell the resources at a guaranteed
price. To find the price of this option, we first establish the following corollary.
All proofs are deferred to the appendix.

\begin{cor*}
\textbf{\emph{(\ref{cor:future-zero})}} The price of a future to buy shares
of the resources on the cheapest path between two network nodes at a \( T_{1} \)
that are resold at \( T_{2} \) is zero.
\end{cor*}
However, to balance the load, the price of the derivative must depend on the
resource prices, therefore the so-called bundle future above is not suitable
for load-balancing. Instead, we define a network option, in the following way.

\begin{defn*}
A network call-option gives the holder the right to send packets with a specified
intensity through nodes on a path between two network routers from time \( T_{1} \)
to time \( T_{2} \), if the fee \( K \) is paid at \( T_{1} \).
\end{defn*}
The call-option price depends on the price of the shares in all routers that
are on any of the possible paths. The following is a very useful theorem for
deriving prices of options based on the correlated price processes. 

\begin{thm*}
\textbf{\emph{(\ref{th:multi-girsanov})}} Let \( \mathbf{S}(T)=\{S_{1}(T),...,(S_{N}(T)\} \)
be an \( N \)-dimensional lognormal price process with correlation \( \rho _{ij}=Corr[dW_{i},dW_{j}] \)
under probability measure \( Q \). Then \[
E^{Q}\left[ S_{m}(T)g(\mathbf{S}(T))|F_{0}\right] =S_{m,0}e^{rT}E^{Q}\left[ g(\xi _{m1}^{T}S_{1}(T),...,\xi _{mN}S_{N}(T))|F_{0}\right] \]
where \[
\xi _{mi}=e^{\sigma _{i}\sigma _{m}\rho _{im}}=exp\left( \frac{1}{dt}\textrm{Cov}\left[ \textrm{log }dS_{i}(T),\textrm{log }dS_{m}(T)\right] \right) \]

\end{thm*}
This shows that linear derivatives can be priced as expected values of an adjusted
process \( \xi _{mi}S_{i}(T) \). 

With the help of this theorem, we can derive the value of the network call-option,
and its partial derivatives, and calculate the optimal rebalance strategy for
a portfolio for the continuous time hedge strategy, using Eq. (\ref{gamma})
and (\ref{beta}).

\begin{cor*}
\textbf{\emph{(\ref{cor:net-price})}} The value of a network call-option with
strike price \( K \) is \begin{eqnarray*}
f(0,\bar{S}) & = & TC\, e^{rT_{1}}\sum _{m=1}^{N}S_{m,0}\sum _{i=1}^{M}v_{im}Q\left[ i=argmin_{j}\hat{C}_{jm}\wedge \hat{C}_{im}<K\right] \\
 &  & \qquad -TC\, K\, Q\left[ \textrm{min}_{j}C_{j}>K\right] 
\end{eqnarray*}
where \[
\hat{C}_{im}=\sum _{k}v_{ik}\xi _{mi}^{T_{1}}S_{k}(T_{1})\]
 is the adjusted cost of path \( i \), after the Girsanov transform to eliminate
resource \( S_{m}(...) \), \[
TC=\frac{e^{-rT_{1}}-e^{-rT_{2}}}{r}\]
 with \( lim_{r\rightarrow 0}TC=T_{2}-T_{1} \).
\end{cor*}
~

\begin{cor*}
\textbf{\emph{(}}\textbf{\ref{cor:net-deriv}}\textbf{\emph{)}} The partial
derivative of the network option with strike-price \( K \) is\[
\frac{\partial f}{\partial S_{n,0}}(0,\bar{S})=TC\, e^{rT_{1}}\sum _{i=1}^{M}v_{in}Q[\hat{C}_{in}=min_{j}\hat{C}_{jn}\wedge \hat{C}_{in}>K]\]
and \[
f(0,\bar{S})=\sum _{m=1}^{N}S_{m,0}\frac{\partial f}{\partial S_{n,0}}(0,\bar{S})-TC\, K\, Q\left[ \textrm{min}_{j}C_{j}>K\right] \]

\end{cor*}
There is no closed form for the sum of lognormal variables \cite{milevsky98asian},
which makes it difficult to reduce the \( Q[...] \)-terms further, but since
\( S(T) \) has a closed form under the risk neutral measure \( Q \), the option
price can be approximated with Monte-Carlo simulation. Note that under the risk
neutral measure one, \( S(T) \) can be simulated without having to simulate
the individual price trajectories for the mean-reverting process, something
that is required for pricing techniques using the natural measure, and which
is very time consuming.

\section{Discussion\label{sec:discussion}}

In the proposed network model, access to each node is traded in a different
market. This way, applications at the edge of the network can combine the resources
in which way they choose, i.e. build broadcast trees, or failure safe multi-path
routing. By choosing to trade capacity shares rather than time-slotted access
as the fundamental commodity, we have only one market per router instead of
one per router and minute.

In the most popular alternative model, network access is handled only at the
edge of the network, hence applications at the edges cannot create new services.
The cost of a more fine grained control scheme is of course more overhead, but
with the proposed scheme, the routers are relieved of much work, since the packets
are source routed. 

The central idea is that simple resources are exchanged on very fast markets,
and that all more complex services are expressed as derivative contract on these
resources. Since we use Farmer markets, the trading generates very little overhead
but incurs a price risk for the trader, which must be hedged. Since end-users
do not hedge their risks themselves, but instead buy derivatives, the network
will not be flooded by bids and quotes between all end-users and all markets.
When a user requires a service, or a combination of services, the user simply
tells a middle-man that it is willing to pay a certain amount for a derivative
that models the service, and the user can be informed directly whether or not
it got the service, and of the marginal cost. 

The capacity prices are assumed to be correlated It\^o processes with multiplicative
drift. We show how use a Martingale technique to price options on one-of-several
linear combinations of assets by proving a theorem on a Girsanov transform that
can be used to price several other options, and give a continuous time hedge
strategy that can be used by a trader to balance out all risk. We have not found
a complete adjusted strategy for the mean-reverting price process when the portfolio
is infrequently rebalanced. A potential possibility to find an adjusted strategy
is to look into pricing of Bessel processes \cite{geman93bessel}. 

The proposed hedge scheme builds on the assumptions made in the Black-Scholes
model, i.e. that the portfolio is self-financing, without arbitrage possibilities,
and that short selling is allowed. Other hedge schemes use different assumptions,
such as CAPM \cite{bode96investments}, which aims to minimize the variance
of the portfolio value while maximizing its expected value. The proposed scheme
has the advantage that the price is invariant of risk-attitude, and can be efficiently
evaluated using Monte-Carlo techniques.

Future work will consist of simulations of the complete bandwidth market in
order to determine the effect of the network options on the price processes,
to find better models for the financial risk of trading in Farmer markets, and
to model other network services as derivative services.

\begin{acknowledgement*}
The author wishes to thank Erik Aurell for advice and many constructive comments
on the topic of financial mathematics, and Sverker Janson for helpful discussions
on agent based models. This work is funded in part by Vinnova/Nutek, The Swedish
Agency for Innovation Systems, program PROMODIS, and in part by SITI, The Swedish
Institute for Information Technology, program Internet3.
\end{acknowledgement*}

\pagebreak

\appendix

\section{~}

Here we give the proofs of the theorems and lemmas needed for pricing the network
option.

\subsection{Bundle futures}

\begin{defn}
A bundle future gives the holder a set of resource shares between \( T_{1} \)
and \( T_{2} \), given that an event \( R \) occurs at \( T_{1} \). Let \( Q \)
be the equivalent martingale measure.
\end{defn}
\begin{thm}
\label{th:bundle}The price of the bundle future is zero.
\begin{lem}
If W(T) is a Wiener process, then \( E\left[ e^{(-\frac{\sigma ^{2}}{2})T+\sigma W(T)}|F_{0}\right] =1 \),
where \( F_{0} \) is the natural filtration up to \( t=0 \). 
\end{lem}
\end{thm}
\begin{lem}
\label{lem:one-res}The price of a future to buy a single resource \( S_{j} \)
at \( T_{1} \) that is sold at \( T_{2} \) is zero.
\end{lem}
\textbf{Proof:} At \( T_{2} \), the future is worth \( A_{j}=S_{j}(T_{2})-e^{r(T_{2}-T_{1})}S_{j}(T_{1}) \).
Hence the price at \( t=0 \) is \begin{eqnarray*}
f(0,s_{0}) & = & e^{-rT_{2}}E^{Q}[A_{j}|F_{0}]\\
 & = & e^{-rT_{2}}\, s_{0}\, E^{Q}\left[ e^{(r-\frac{\sigma ^{2}}{2})T_{2}+\sigma W(T_{2})}-e^{r(T_{2}-T_{1})}e^{(r-\frac{\sigma ^{2}}{2})T_{1}+\sigma W(T_{1})}|F_{0}\right] \\
 & = & s_{0}\, E^{Q}\left[ e^{(-\frac{\sigma ^{2}}{2})T_{2}+\sigma W(T_{2})}-e^{(-\frac{\sigma ^{2}}{2})T_{1}+\sigma W(T_{1})}|F_{0}\right] \\
 & = & s_{0}\, E^{Q}\left[ e^{(-\frac{\sigma ^{2}}{2})T_{1}+\sigma W(T_{1})}|F_{0}\right] \underbrace{E^{Q}\left[ e^{(-\frac{\sigma ^{2}}{2})(T_{2}-T_{1})+\sigma (W(T_{2})-W(T_{1}))}-1|F_{0}\right] }_{=0}\\
 & = & 0
\end{eqnarray*}
\qed

\begin{lem}
The price of a derivative that delivers the future defined in lemma \ref{lem:one-res},
given that event \( R \) occurs at \( T_{1} \), is also zero.
\end{lem}
\textbf{Proof:} At \( T_{2} \) the derivative is worth \( A_{j}1_{\{R\}} \).
Hence the option price at \( t=0 \) is\begin{eqnarray*}
f(0,s_{0}) & = & e^{rT_{2}}E^{Q}[A_{j}1_{\{R\}}|F_{0}]\\
 & = & e^{rT_{2}}E^{Q}\left[ \underbrace{E^{Q}[A_{j}|F_{1}]}_{=0}Q[R]|F_{0}\right] \\
 & = & 0
\end{eqnarray*}
where \( Q[R] \) is the probability of \( R \) under the probability measure
\( Q \). \qed

\textbf{Proof} \textbf{of Theorem} \ref{th:bundle}\textbf{:} Let \( R_{i} \)
be the event that bundle \( i \) is bought. The price of the bundle future
is\begin{eqnarray*}
f(0,\bar{S}) & = & e^{-rT_{2}}E^{Q}\left[ \sum ^{N}_{j=1}v_{ij}A_{j}1_{\{R_{i}\}}|F_{0}\right] \\
 & = & 0
\end{eqnarray*}
\qed

\begin{cor}
\label{cor:future-zero}The price of a future to buy shares of the resources
on the cheapest path between two network nodes at a \( T_{1} \) that are resold
at \( T_{2} \) is zero.
\end{cor}
\textbf{Proof:} Let \( v_{ij} \) be the amount of router \( j \) required
on path \( i \), and let \( R_{i} \) be the event that path \( i \) is the
cheapest path at \( T_{1} \). The corollary follows from Theorem \ref{th:bundle}. \qed

\subsection{Network option (step one)}

\begin{defn}
A network option gives the holder the right to send packets with a specified
intensity through nodes on a path between two network routers from time \( T_{1} \)
to time \( T_{2} \), if the fee \( K \) is paid at \( T_{1} \).
\end{defn}
\begin{lem}
The price of an arithmetic average (Asian) call option with strike price zero
and maturity at \( T \) is\[
f(0,s_{0})=s_{0}\frac{1-e^{-rT}}{r}\]
which becomes \( s_{0}T \) in the limit of \( r\rightarrow 0^{+} \). 
\end{lem}
\textbf{Proof:} At \( T \), the option is worth \( \int _{0}^{T}S(t)dt \),
so\begin{eqnarray*}
f_{\textrm{Asian}}(0,s_{0}) & = & e^{-rT}E^{Q}\left[ \int _{0}^{T}S(t)dt|F_{0}\right] \\
 & = & \textrm{lim}_{\Delta t\rightarrow 0}e^{-rT}E^{Q}\left[ \sum _{i=0}^{T/\Delta t-1}s_{0}e^{(r-\frac{\sigma ^{2}}{2})t_{i}+\sigma W(t_{i})}\Delta t|F_{0}\right] \\
 & = & \textrm{lim}_{\Delta t\rightarrow 0}e^{-rT}\sum _{i=0}^{T/\Delta t-1}s_{0}e^{r\, t_{i}}\Delta t\, \underbrace{E^{Q}\left[ e^{(-\frac{\sigma ^{2}}{2})t_{i}+\sigma W(t_{i})}|F_{0}\right] }_{=1}\\
 & = & e^{-rT}\int _{0}^{T}s_{0}e^{rt}dt\\
 & = & s_{0}\frac{1-e^{-rT}}{r}
\end{eqnarray*}
\qed

\begin{thm}
Let \( C_{i}=\sum _{m=1}^{N}v_{im}S_{j}(T_{1}) \) be the cost of the resources
for alternative \( i \) at \( T_{1} \). The price of a network option is\begin{eqnarray*}
f(0,\bar{S}) & = & TC\, E^{Q}\left[ max(min_{i}(C_{i})-K,0)|F_{0}\right] \\
 & = & TC\left( E^{Q}\left[ \sum _{i=1}^{M}C_{i}1_{\{C_{i}=min_{k}C_{k}\}}1_{\{C_{i}>K\}}|F_{0}\right] -K\, E^{Q}\left[ 1_{\{min_{i}C_{i}>K\}}|F_{0}\right] \right) 
\end{eqnarray*}
where \( TC=\frac{e^{-rT_{1}}-e^{-rT_{2}}}{r} \), for which \( \textrm{lim}_{r\rightarrow 0}TC=T_{2}-T_{1} \). 
\end{thm}
\textbf{Proof:} The cost to send traffic consists of buying the required router
shares at \( T_{1} \), pay the send fee from \( T_{1} \) to \( T_{2} \) and
sell back the shares at \( T_{2} \). This amounts to a bundle option and an
Asian option from \( T_{1} \) to \( T_{2} \) on some path between the source
and the destination. The cheapest option is the option over the least cost path,
i.e. where \( i=argmin_{k}C_{k} \) at \( T_{1} \). At \( T_{1} \) the network
option is worth the sum of Asian options for the resources on the cheapest path
minus \( K \) if the option is in-the-money, else it is worth 0. Hence, on
\( t=0 \)\begin{eqnarray*}
f(0,\bar{S}) & = & e^{-rT_{1}}E^{Q}\left[ max\left( \sum _{i=1}^{M}\sum _{m=1}^{N}v_{im}f_{\textrm{Asian}}(T_{1},S(T_{1}))1_{\{C_{i}=min_{k}C_{k}\}}-K,0\right) |F_{0}\right] \\
 & = & e^{-rT_{1}}E^{Q}\left[ max\left( \sum _{i=1}^{M}\frac{1-e^{-r(T_{2}-T_{1})}}{r}\sum _{m=1}^{N}v_{im}S(T_{1})1_{\{C_{i}=min_{k}C_{k}\}}-K,0\right) |F_{0}\right] \\
 & = & \frac{e^{-rT_{1}}-e^{-rT_{2}}}{r}E^{Q}\left[ max(\sum _{i=1}^{M}C_{i}1_{\{C_{i}=min_{k}C_{k}\}}-K,0)|F_{0}\right] \\
 & = & TC\, \left( E^{Q}\left[ \sum _{i=1}^{M}C_{i}1_{\{C_{i}=min_{k}C_{k}\}}1_{\{C_{i}>K\}}|F_{0}\right] -K\, E^{Q}\left[ 1_{\{min_{i}C_{i}>K\}}|F_{0}\right] \right) 
\end{eqnarray*}
\qed

\subsection{The 1-dimensional Girsanov transform}

We start by showing the usefulness of the so-called Girsanov transform for a
one-dimensional stochastic process, in order to simplify the presentation of
the proof of the \( n \)-dimensional transform. The one-dimensional pricing
of a call option was based on Dufresne \emph{et al}. \cite{dufresnte96pricing}.

\begin{lem}
\( E^{Q}\left[ S(T)g(S(T))|F_{0}\right] =S_{0}e^{rT}E^{Q}\left[ g(e^{\sigma ^{2}T}S(T))|F_{0}\right]  \)
\end{lem}
\textbf{Proof:} \begin{eqnarray*}
E^{Q}[S(T)g(S(T))|F_{0}] & = & \int _{-\infty }^{\infty }S_{0}e^{(r-\frac{\sigma ^{2}}{2})T+\sigma \sqrt{T}x}\frac{1}{\sqrt{2\pi }}e^{-\frac{1}{2}x^{2}}g\big (S_{0}e^{(r-\frac{\sigma ^{2}}{2})T+\sigma \sqrt{T}x}\big )dx\\
 & = & S_{0}e^{rT}\int _{-\infty }^{\infty }\frac{1}{\sqrt{2\pi }}e^{-\frac{1}{2}(x-\sigma \sqrt{T})^{2}}g\big (S_{0}e^{(r-\frac{\sigma ^{2}}{2})T+\sigma \sqrt{T}x}\big )dx\\
 & = & S_{0}e^{rT}\int _{-\infty }^{\infty }\frac{1}{\sqrt{2\pi }}e^{-\frac{1}{2}y^{2}}g\big (S_{0}e^{(r-\frac{\sigma ^{2}}{2})T+\sigma \sqrt{T}(y+\sigma \sqrt{T})}\big )dx\\
 & = & S_{0}e^{rT}E^{Q}\left[ g(e^{\sigma ^{2}T}S(T))|F_{0}\right] 
\end{eqnarray*}
where \( y=x-\sigma \sqrt{T} \). \qed

\begin{cor}
The value of a call option with strike price \( K \), and maturity at \( T \),
is \[
e^{-rT}E^{Q}\left[ \textrm{max}(S(T)-K,0)|F_{0}\right] =S_{0}N(d_{1}+\sigma \sqrt{T})-Ke^{-rT}N(d_{1})\]
where \( d_{1}=\frac{\textrm{log}\frac{S_{0}}{K}-\frac{\sigma ^{2}}{2}T}{\sigma \sqrt{T}} \),
\( N(x) \) is the cumulative density function for the standard normal distribution,
the current time \( t=0, \) and \( S(0)=S_{0} \).
\end{cor}
\textbf{Proof:} From the definition of the indicator function, \( E^{Q}[1_{\{A\}}]=Q[A] \).
The corollary follows from seeing that \[
E^{Q}[\textrm{max}(S(T)-K,0)|F_{0}]=E^{Q}[S(T)\, 1_{\{S(T)>K\}}|F_{0}]-K\, E^{Q}[1_{\{S(T)>K\}},0)|F_{0}]\]
and from the lemma, \begin{eqnarray*}
E^{Q}[1_{\{S(T)>K\}}|F_{0}] & = & Q\Big [-\frac{W(T)}{\sqrt{T}}<\underbrace{\frac{\textrm{log}\frac{S_{0}}{K}-\frac{\sigma ^{2}}{2}T}{\sigma \sqrt{T}}}_{=d_{1}}\Big ]\\
 & = & N(d_{1})
\end{eqnarray*}
and \begin{eqnarray*}
E^{Q}[S(T)\, 1_{\{S(T)>K\}}|F_{0}] & = & S_{0}e^{rT}E^{Q}\left[ 1_{\{e^{\sigma ^{2}T}S(T)>K\}}|F_{0}\right] \\
 & = & S_{0}e^{rT}Q\left[ -\frac{W(T)}{\sqrt{T}}<\frac{\textrm{log}\frac{S_{0}}{K}-\frac{\sigma ^{2}}{2}T}{\sigma \sqrt{T}}+\sigma \sqrt{T}\right] \\
 & = & S_{0}e^{rT}N(d_{1}+\sigma \sqrt{T})
\end{eqnarray*}
 by doing a Girsanov transform, and using that \( W(T) \) is normal distributed
with variance \( T \). \qed

\subsection{The \protect\( n\protect \)-dimensional Girsanov transform of \protect\( E[\mathbf{S}\, g(\mathbf{S})]\protect \)}

\begin{lem}
\label{lem:uncorr}Let \( \{\mathbf{S}^{Q}(T)\}_{i}=S_{i,0}e^{(r-\frac{\sigma _{i}^{2}}{2})T+\sigma _{i}\sqrt{T}x_{i}} \),
\( i=i,...,N \), where \( x_{i} \) are standard normal distributed variables,
correlated with \( \{D\}_{ij}=\textrm{Corr}[x_{i},x_{j}] \) under the measure
\( Q \), and uncorrelated under the measure \( R \). Then \[
E^{Q}[g(\mathbf{S}^{Q}(T))|F_{0}]=E^{R}[g(\mathbf{S}^{R}(T))|F_{0}]\]
where \( \{\mathbf{S}^{R}(T)\}_{i}=S_{i,0}e^{(r-\frac{\sigma _{i}^{2}}{2})T+\sigma _{i}\sqrt{T}\sum _{j=1}^{N}P_{ij}x_{j}} \),
and \( P \) is the Cholesky factorization of \( D \), i.e. \( P^{T}P=D \).
\end{lem}
\textbf{Proof:} Since \( P^{T}P=D \), then \( D^{-1}P^{T}P=I=P^{T}D^{-1}P \).
The substitution \( P\mathbf{y}=\mathbf{x} \) makes \( \mathbf{x}^{T}D^{-1}\mathbf{x}=\mathbf{y}^{T}\mathbf{y} \),
\( d\mathbf{x}=(\textrm{det }P)d\mathbf{y}=\sqrt{\textrm{det }D}d\mathbf{y} \),
and \( x_{i}=\sum _{j=1}^{N}P_{ij}y_{j} \).

\begin{eqnarray*}
E^{Q}[g(\mathbf{S}(T))|F_{0}] & = & \int _{-\infty }^{\infty }d\mathbf{x}\, g(\mathbf{S}^{Q}(T))\frac{1}{(2\pi )^{N/2}\sqrt{\textrm{det }D}}e^{-\frac{1}{2}\mathbf{x}^{T}D^{-1}\mathbf{x}}\\
 & = & \int _{-\infty }^{\infty }d\mathbf{y}\, g(\mathbf{S}^{R}(T))\frac{1}{(2\pi )^{N/2}}e^{-\frac{1}{2}\mathbf{y}^{T}\mathbf{y}}\\
 & = & E^{R}[g(\mathbf{S}^{R}(T))|F_{0}]
\end{eqnarray*}
\qed

\begin{lem}
\label{lem:elim}The multidimensional variant of the elimination of \( S_{m} \)
for an uncorrelated N-dimensional random variable \( \mathbf{S} \) is\[
E^{R}[\{\mathbf{S}^{R}(T)\}_{m}g(\mathbf{S}^{R}(T))|F_{0}]=S_{m,0}e^{rT}E^{R}[g(\mathbf{S}^{Z}(T))|F_{0}]\]
 where \( \{\mathbf{S}^{Z}(T)\}_{i}=e^{\sigma _{i}\sigma _{m}\sum _{j=1}^{N}P_{ij}P_{mj}}\{\mathbf{S}^{R}(T)\}_{i} \).
\end{lem}
\textbf{Proof:} \( E^{R}[\{\mathbf{S}^{R}(T)\}_{m}g(\mathbf{S}^{R}(T))|F_{0}] \)
\begin{eqnarray*}
 & = & \int _{-\infty }^{\infty }d\mathbf{x}S_{m,0}e^{(r-\frac{\sigma _{m}^{2}}{2})T+\sigma _{m}\sqrt{T}\sum _{j=1}^{N}P_{mj}x_{j}}\frac{1}{(2\pi )^{N/2}}e^{-\frac{1}{2}\sum _{j=1}^{n}x_{j}^{2}}g(\mathbf{S}^{R}(T))\\
 & = & S_{m,0}e^{(r-\frac{\sigma _{m}^{2}}{2})T}\int _{-\infty }^{\infty }d\mathbf{x}\frac{1}{(2\pi )^{N/2}}e^{-\frac{1}{2}\sum _{j=1}^{n}(x_{j}^{2}-2\sigma _{m}\sqrt{T}P_{mj}x_{j})}g(\mathbf{S}^{R}(T))\\
 & = & S_{m,0}e^{(r-\frac{\sigma _{m}^{2}}{2})T+\frac{1}{2}\sigma _{m}^{2}T\overbrace{\sum _{j=1}^{N}P_{mj}}^{=1}}\int _{-\infty }^{\infty }d\mathbf{x}\frac{1}{(2\pi )^{N/2}}e^{-\frac{1}{2}\sum _{j=1}^{n}(x_{j}-\sigma _{m}\sqrt{T}P_{mj})^{2}}g(\mathbf{S}^{R}(T))\\
 & = & S_{m,0}e^{rT}\int _{-\infty }^{\infty }d\mathbf{y}\frac{1}{q(2\pi )^{N/2}}e^{-\frac{1}{2}\mathbf{y}^{T}\mathbf{y}}g(\mathbf{S}^{Z}(T))\\
 & = & S_{m,0}e^{rT}E^{R}[g(\mathbf{S}^{Z}(T))|F_{0}]
\end{eqnarray*}
 where \( x_{i}=y_{i}+\sigma _{m}\sqrt{T}P_{mi} \), therefore \begin{eqnarray*}
\{\mathbf{S}^{Z}(T)\}_{i} & = & S_{i,0}e^{(r-\frac{\sigma _{i}^{2}}{2})T+\sigma _{i}\sqrt{T}\sum _{j=1}^{N}P_{ij}(y_{j}+\sigma _{m}\sqrt{T}P_{mj})}\\
 & = & e^{\sigma _{i}\sigma _{m}T\sum _{j=1}^{N}P_{ij}P_{mj}}S_{i,0}e^{(r-\frac{\sigma _{i}^{2}}{2})T+\sigma _{i}\sqrt{T}\sum _{j=1}^{N}P_{ij}y_{j})}\\
 & = & e^{\sigma _{i}\sigma _{m}T\sum _{j=1}^{N}P_{ij}P_{mj}}\{\mathbf{S}^{R}(T)\}_{i}\\
 & = & e^{\sigma _{i}\sigma _{m}T\{D\}_{im}}\{\mathbf{S}^{R}(T)\}_{i}
\end{eqnarray*}
\qed

\begin{lem}
\( \sigma _{i}\sigma _{m}\{D\}_{im}=\frac{1}{dt}\textrm{Cov}^{Q}[\textrm{log }dS_{i}(T),\textrm{log }dS_{m}(T)] \).
\end{lem}
\textbf{Proof:} By It\^o's Formula (see for instance, theorem 3.3.2 in \cite{kloeden99numerical}) \( log\, dS_{i}=a\, dt+\sigma _{i}dW_{i} \)
for some bounded function \( a \). \begin{eqnarray*}
\textrm{Cov}^{Q}[\textrm{log }dS_{i}(T),\textrm{log }dS_{m}(T)] & = & \textrm{Cov}^{Q}[\sigma _{i}dW_{i},\sigma _{j}dW_{j}]+O(dt^{3/2})\\
 & = & \sigma _{i}\sigma _{j}\{D\}_{ij}dt+O(dt^{3/2})
\end{eqnarray*}
\qed

\begin{thm}
\label{th:multi-girsanov}Let \( \mathbf{S}(T)=\{S_{1}(T),...,(S_{N}(T)\} \)
be an \( N \)-dimensional lognormal price process with \emph{correlation \( \{D\}_{ij}=Corr[dW_{i},dW_{j}] \)}
under probability measure \( Q \). Then \begin{eqnarray*}
E^{Q}[S_{m}(T)g(\mathbf{S}(T))|F_{0}] & = & S_{m,0}e^{rT}E^{Q}[g(\xi _{m1}^{T}S_{1}(T),...,\xi _{mN}S_{N}(T))|F_{0}]\\
 &  & 
\end{eqnarray*}
where \( \xi _{mi}=e^{\sigma _{i}\sigma _{m}\{D\}_{im}}=exp(\frac{1}{dt}\textrm{Cov}^{Q}[\textrm{log }dS_{i}(T),\textrm{log }dS_{m}(T)]) \).
\end{thm}
\textbf{Proof:} Follows from using lemma \ref{lem:uncorr}, lemma \ref{lem:elim},
and lemma \ref{lem:uncorr} again, in the other direction.\qed

\subsection{Network option (step two)}

\begin{cor}
\label{cor:net-price}The value of a network call-option with strike price \( K \)
is \begin{eqnarray*}
f(0,\bar{S}) & = & TC\, e^{rT_{1}}\sum _{m=1}^{N}S_{m,0}\sum _{i=1}^{M}v_{im}Q[i=argmin_{j}\hat{C}_{jm}\wedge \hat{C}_{im}>K]\\
 &  & \qquad -TC\, K\, Q[\textrm{min}_{j}C_{j}>K]
\end{eqnarray*}
where \( \hat{C}_{im}=\sum _{k}v_{ik}\xi _{mi}^{T_{1}}S_{k}(T_{1}) \) is the
adjusted cost of path \( i \), after the Girsanov-transform to eliminate resource
\( S_{m}(...) \).
\end{cor}
\textbf{Proof:} \[
f(0,\bar{S})=TC\, \Big (\underbrace{E^{Q}\left[ \sum _{i=1}^{M}C_{i}1_{\{C_{i}=min_{k}C_{k}\}}1_{\{C_{i}>K\}}|F_{0}\right] }_{=V_{1}}-\underbrace{K\, E^{Q}[1_{\{min_{i}C_{i}>K\}}|F_{0}]}_{=V_{2}}\Big )\]
where \( V_{2}=K\, Q[min_{i}C_{i}>K] \) and \begin{eqnarray*}
V_{1} & = & E^{Q}[\sum _{i=1}^{M}C_{i}1_{\{C_{i}=min_{k}C_{k}\}}1_{\{C_{i}>K\}}|F_{0}]\\
 & = & E^{Q}[\sum _{i=1}^{M}\sum _{m=1}^{N}v_{im}S_{m}(T_{1})1_{\{C_{i}=min_{k}C_{k}\wedge C_{i}>K\}}|F_{0}]\\
 & = & \sum _{i=1}^{M}\sum _{m=1}^{N}v_{im}S_{0.m}e^{rT_{1}}E^{Q}[1_{\{\hat{C}_{im}=min_{k}\hat{C}_{km}\wedge \hat{C}_{im}>K\}}|F_{0}]\\
 & = & e^{rT_{1}}\sum _{m=1}^{N}S_{0.m}\sum _{i=1}^{M}v_{im}Q[\hat{C}_{im}=min_{k}\hat{C}_{km}\wedge \hat{C}_{im}>K]
\end{eqnarray*}
by using theorem \ref{th:multi-girsanov} in step three.\qed

\begin{cor}
\label{cor:net-deriv}The partial derivative of the network option with strike-price
\( K \) is\[
\frac{\partial f}{\partial S_{n,0}}(0,\bar{S})=TC\, e^{rT_{1}}\sum _{i=1}^{M}v_{in}Q[\hat{C}_{in}=min_{j}\hat{C}_{jn}\wedge \hat{C}_{in}>K]\]
and \[
f(0,\bar{S})=\sum _{m=1}^{N}S_{m,0}\frac{\partial f}{\partial S_{n,0}}(0,\bar{S})-TC\, K\, Q[\textrm{min}_{j}C_{j}>K]\]

\end{cor}
\textbf{Proof:} The first statement follows from that \( \frac{\partial Q^{*}}{\partial S_{in}}=0 \)
nearly everywhere, for all \( m \) and \( n \), for both cases when \( Q^{*}=Q[\hat{C}_{in}=min_{j}\hat{C}_{jn}\wedge \hat{C}_{in}>K] \),
and when \( Q^{*}=Q[\textrm{min}_{j}C_{j}>K] \). The second follows trivially
from the value of the network call-option and the first statement.\qed
\end{document}